%% file: main.tex
\pdfoutput=1
\documentclass[11pt]{article}
\usepackage[final]{acl}
\usepackage{times}
\usepackage{latexsym}
\usepackage[T1]{fontenc}
\usepackage[utf8]{inputenc}
\usepackage{hyperref}
\usepackage{microtype}
\usepackage{colortbl}
\usepackage{inconsolata}
\usepackage{graphicx}
\usepackage{amsfonts,amsmath,amssymb}
\usepackage{booktabs}
\usepackage{arydshln}
\usepackage[table]{xcolor}
\usepackage{booktabs,tabularx,array,xcolor}
\usepackage{tcolorbox}
\tcbuselibrary{breakable}
\newcolumntype{Y}{>{\raggedright\arraybackslash}X}

\definecolor{goodcell}{RGB}{209, 229, 240}    
\definecolor{badcell}{RGB}{253, 208, 162}     

\colorlet{tableheadcolor}{gray!25} 
\colorlet{tablerowcolor}{gray!10} 
\colorlet{tablerowcolor2}{gray!45} 
\colorlet{tablerowcolor3}{gray!12} 

\newcommand{\rowcollight}{\rowcolor{tablerowcolor3}} %

\title{Masking or Mitigating?\\Deconstructing the Impact of Query Rewriting on Retriever Biases in RAG}

\author{
 \textbf{Agam Goyal\textsuperscript{$\spadesuit$,$\ast$,$\ddagger$}},
 \textbf{Koyel Mukherjee\textsuperscript{$\clubsuit$,$\ddagger$}},
 \textbf{Apoorv Saxena\textsuperscript{$\heartsuit$}},
 \textbf{Anirudh Phukan\textsuperscript{$\diamondsuit$}},
\\
 \textbf{Eshwar Chandrasekharan\textsuperscript{$\spadesuit$},}
 \textbf{Hari Sundaram\textsuperscript{$\spadesuit$}}
\\
 \textsuperscript{$\spadesuit$}Siebel School of Computing and Data Science,  University of Illinois Urbana-Champaign\\
 \textsuperscript{$\clubsuit$}Adobe Research\qquad
 \textsuperscript{$\heartsuit$}Inception Labs\qquad
 \textsuperscript{$\diamondsuit$}Indian Institute of Science
\\
\texttt{\{agamg2, eshwar, hs1\}@illinois.edu, komukher@adobe.com}
}

\begin{document}
\maketitle
\def\thefootnote{$\ast$}\footnotetext{Work done during internship at Adobe Research}
\def\thefootnote{$\ddagger$}\footnotetext{Corresponding Authors}

\begin{abstract}
Dense retrievers in retrieval-augmented generation (RAG) systems exhibit systematic biases---including brevity, position, literal matching, and repetition biases---that can compromise retrieval quality. Query rewriting techniques are now standard in RAG pipelines, yet their impact on these biases remains unexplored. We present the first systematic study of how query enhancement techniques affect dense retrieval biases, evaluating five methods across six retrievers. Our findings reveal that simple LLM-based rewriting achieves the strongest aggregate bias reduction (54\%), yet fails under adversarial conditions where multiple biases combine. Mechanistic analysis uncovers two distinct mechanisms: simple rewriting reduces bias through increased score variance, while pseudo-document generation methods achieve reduction through genuine decorrelation from bias-inducing features. However, no technique uniformly addresses all biases, and effects vary substantially across retrievers. Our results provide practical guidance for selecting query enhancement strategies based on specific bias vulnerabilities. More broadly, we establish a taxonomy distinguishing query-document interaction biases from document encoding biases, clarifying the limits of query-side interventions for debiasing RAG systems.
\end{abstract}

\input{latex/1Introduction}
\input{latex/2RelatedWorks}
\input{latex/3Methods}
\input{latex/4Results}
\input{latex/5Discussion}
\input{latex/6Conclusion}
\input{latex/7Limitations}
\input{latex/Acknowledgments}
\input{latex/EthicsStatement}

\bibliography{references}

\appendix
\input{latex/Appendix}

\end{document}

%% file: latex/1Introduction.tex
\section{Introduction}

Retrieval-Augmented Generation (RAG) systems have become foundational for grounding large language models in external knowledge, enabling applications from question-answering to document analysis across diverse domains~\cite{lewis2020retrieval,lu-etal-2022-reacc,jiang2023hykge,gao2023retrieval,asai2024reliable}. These systems typically employ dense retrievers~\cite{karpukhin-etal-2020-dense,thakur2021beir} to identify relevant passages from document collections, which are then provided as context for response generation~\cite{oche2025systematic}.


\begin{figure}
    \centering
    \includegraphics[width=0.98\linewidth]{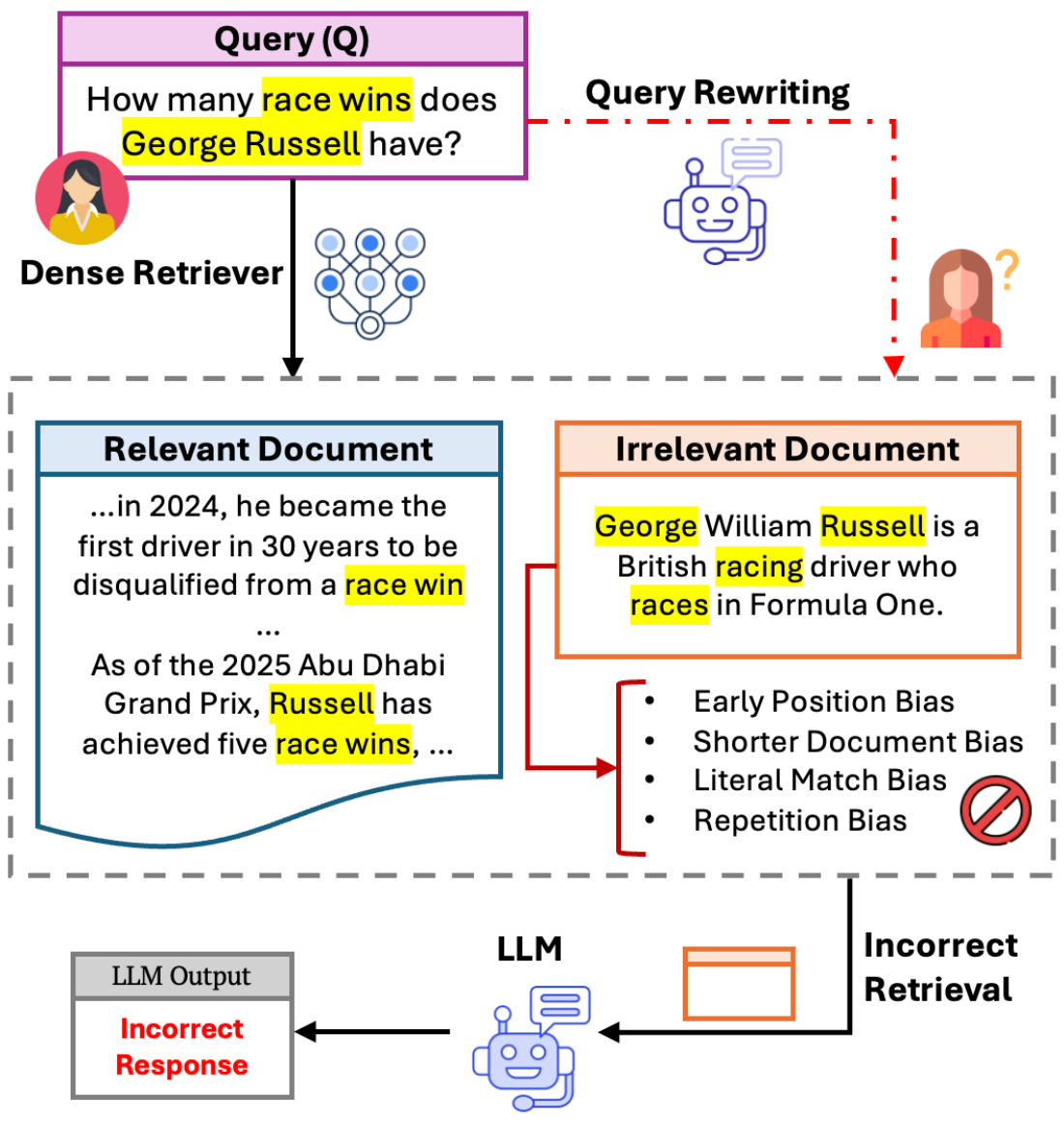}
    \caption{\textbf{Paper Overview.} Dense retrievers are susceptible to various biases, leading to incorrect retrieval and erroneous LLM outputs. Can query rewriting help? We systematically evaluate its effects on retrieval bias.}
    \label{fig:teaser-figure}
\end{figure}

Despite their effectiveness, recent work has revealed that dense retrievers exhibit systematic biases that can compromise retrieval quality~\cite{ram-etal-2023-token,coelho-etal-2024-dwell}. \citet{fayyaz-etal-2025-collapse} demonstrated that retrievers consistently favor documents with surface-level characteristics---brevity, early answer positioning, lexical overlap, and entity repetition---over documents containing actual factual evidence. The ideal solution would be dense retrieval models inherently robust to such biases and capable of prioritizing semantic relevance. However, achieving this through architectural modifications or training objective changes remains challenging. A more tractable first step is understanding whether these biases persist under query enhancement techniques~\cite{ma-etal-2023-query}, which are standard in modern RAG pipelines.

Production RAG systems routinely employ query transformations including LLM-based rewriting for better semantic matching and pseudo-document generation approaches like HyDE~\cite{gao-etal-2023-precise} and Query2Doc~\cite{wang-etal-2023-query2doc}. While these techniques improve retrieval effectiveness, their interaction with systematic biases remains unexplored: do they amplify existing biases by generating surface-level features that retrievers favor, or do specific strategies mitigate particular bias types? Understanding this interaction is critical for deploying robust and fair RAG systems, particularly in high-stakes domains such as medical, sociotechnical, educational, and enterprise settings~\cite{ryan2025enronqa,wang2025pir,goyal-etal-2025-momoe,shimgekar2025detecting,zhan2025slm,goyal2026vastu,pal2026hidden}.

In this work, \textbf{we conduct the first systematic investigation into how query rewriting techniques affect retrieval biases}. We evaluate five enhancement methods across six retrievers using the controlled framework of \citet{fayyaz-etal-2025-collapse}. This framework's treatment-control methodology, which uses paired documents that differ only in specific surface characteristics while maintaining factual equivalence, enables us to isolate and measure the effect of query rewriting on individual bias types rather than confounding bias reduction with overall retrieval improvements. We find that query enhancement effects are highly differential: simple LLM rewriting achieves the strongest aggregate bias reduction (54\%) but fails under adversarial conditions, while pseudo-document methods provide more modest but mechanistically robust improvements. We establish a taxonomy distinguishing \textit{interaction biases} amenable to query-side interventions from \textit{encoding biases} requiring retriever-level modifications, providing practical guidance for bias-aware RAG deployment.

%% file: latex/2RelatedWorks.tex
\section{Related Works}

\paragraph{Bias in Dense Retrieval.} Dense retrieval has emerged as a powerful alternative to traditional sparse methods like BM25 for information retrieval tasks~\cite{karpukhin-etal-2020-dense,thakur2021beir}, showing strong performance on retrieval benchmarks.  However, despite these advances, recent work has begun to uncover systematic biases in how dense retrievers encode and rank documents, including position biases~\cite{coelho-etal-2024-dwell} and lexical biases~\cite{ram-etal-2023-token}. Similarly, \citet{behnamghader-etal-2023-retriever} show that Dense Passage Retrieval (DPR) models fail to retrieve statements requiring reasoning beyond surface-level similarity. In terms of downstream consequences of these biases, \citet{huang2026answer} and \citet{samuel2026does} show how these biases can affect the coverage, presentation, and fairness of the outputs from generative search systems. Most relevant to our work, \citet{fayyaz-etal-2025-collapse} introduced ColDeR, a controlled benchmark for systematically measuring various types of retrieval biases that biases persist across multiple dense retriever architectures and can compound adversarially. \emph{Our work extends this line of research by investigating whether and how query-side interventions can mitigate these documented biases.}

\paragraph{Query Enhancement in RAG.} RAG systems~\cite{lewis2020retrieval,gao2023retrieval} have become foundational for knowledge-intensive NLP tasks, with query rewriting now a standard pipeline component~\cite{ma-etal-2023-query}. Several query enhancement techniques have been proposed to bridge the semantic gap between queries and documents. HyDE~\cite{gao-etal-2023-precise} generates pseudo-documents and retrieves based on their embeddings rather than the original query; Query2Doc~\cite{wang-etal-2023-query2doc} concatenates LLM-generated pseudo-documents to expand queries; and various other simple LLM-based query rewriting techniques have been proposed in prior work~\cite{jagerman2023query,mao-etal-2024-rafe,kim-etal-2025-gure}. While these methods improve retrieval effectiveness, their impact on systematic retrieval biases remains unexplored. As RAG systems are increasingly deployed in high-stakes domains, understanding how retrieval biases propagate through these pipelines becomes critical. \emph{Our work fills this gap by analyzing how query enhancement techniques interact with known retrieval biases.}

%% file: latex/3Methods.tex
\section{Preliminaries, Data, and Methods}

\subsection{Problem Formulation}

We investigate whether and how query rewriting techniques affect systematic biases in dense retrievers. We adopt the treatment-control ColDeR benchmark contributed by \citet{fayyaz-etal-2025-collapse}, which uses carefully constructed document pairs using the Re-DocRED dataset~\cite{tan-etal-2022-revisiting} that differ only in specific surface characteristics while maintaining factual equivalence.

Given a query $q$ and a pair of documents $(D_1, D_2)$ where $D_1$ contains a bias-inducing characteristic and $D_2$ serves as the control, we measure retrieval preferences using the similarity scores assigned by a dense retriever $\mathcal{S}$.  This controlled design enables us to isolate the effect of query rewriting on individual bias types, separating bias mitigation from general retrieval performance improvements. Then, for each document pair, we compute the difference in retrieval scores: 
\begin{align*} 
\delta = \mathcal{S}(q, D_1) - \mathcal{S}(q, D_2) 
\end{align*} 
where $\mathcal{S}(q, D)$ represents the retriever's similarity score between the generated embeddings for the query $q$ and document $D$. Positive $\delta$ values indicate bias toward surface-level characteristics, while values near zero suggest factual grounding. We apply different query enhancement techniques to transform the original query $q$ into $q'$, and measure how this transformation affects the distribution of $\delta$ across query-document pairs. For each bias type, we collect $N=200$ paired documents $\{(D_1, D_2)\}_i^N$ along with associated queries. We now describe each bias type in detail.




\begin{table*}[t]
\centering
\sffamily
\small
\renewcommand{\arraystretch}{1.3}
\begin{tabular}{@{} p{1.6cm} p{6.8cm} p{6.8cm} @{}}
 & \textbf{Document $D_1$} (Bias-amplified) 
 & \textbf{Document $D_2$} (Control) \\
\midrule
 
\textbf{Brevity Bias}
&
\textbf{Q:} What series is \textbf{Lost Verizon} part of? \newline
\textbf{D:} ``\colorbox{yellow!30}{\textbf{Lost Verizon}}'' is the 
second episode of \colorbox{cyan!20}{\textit{The Simpsons}}' twentieth 
season.
&
\textbf{Q:} What series is \textbf{Lost Verizon} part of? \newline
\textbf{D:} ``\colorbox{yellow!30}{\textbf{Lost Verizon}}'' is the 
second episode of \colorbox{cyan!20}{\textit{The Simpsons}}' twentieth 
season. It first aired on the Fox network in the United States on 
October 5, 2008. Bart becomes jealous of his friends and their cell 
phones\ldots
\\
\midrule
 
\textbf{Literal Bias}
&
\textbf{Q:} When was \textbf{Seyhun} born? \newline
\textbf{D:} \colorbox{yellow!30}{\textbf{Seyhun}}, 
(\colorbox{cyan!20}{August 22, 1920} -- May 26, 2014) was an 
Iranian architect, sculptor, painter, scholar and professor\ldots
&
\textbf{Q:} When was \textbf{Seyhun} born? \newline
\textbf{D:} \colorbox{yellow!30}{Houshang \textbf{Seyhoun}}, 
(\colorbox{cyan!20}{August 22, 1920} -- May 26, 2014) was an 
Iranian architect, sculptor, painter, scholar and professor\ldots
\\
\midrule
 
\textbf{Position Bias}
&
\textbf{Q:} Which country is \textbf{Wonyong Sung} a citizen of? \newline
\textbf{D:} \colorbox{yellow!30}{\textbf{Wonyong Sung}} (born 1950s), 
\colorbox{cyan!20}{\textbf{South Korean}} professor of electronic 
engineering. Won-yong is a Korean masculine given name\ldots
&
\textbf{Q:} Which country is \textbf{Wonyong Sung} a citizen of? \newline
\textbf{D:} Won-yong is a Korean masculine given name\ldots\ 
South Korean swimmer 
\colorbox{yellow!30}{\textbf{Wonyong Sung}} (born 1950s), 
\colorbox{cyan!20}{\textbf{South Korean}} professor of electronic 
engineering
\\
\midrule
 
\textbf{Repetition Bias}
&
\textbf{Q:} Where was \textbf{James Paul Maher} born? \newline
\textbf{D:} Born in \colorbox{cyan!20}{\textbf{Brooklyn, New York}}, 
\colorbox{yellow!30}{\textbf{Maher}} graduated from St.\ Patrick's 
Academy in Brooklyn. \colorbox{yellow!30}{\textbf{James Paul Maher}} 
(Nov.\ 3, 1865 -- Jul.\ 31, 1946) was a U.S.\ Representative 
from New York. \colorbox{yellow!30}{\textbf{Maher}} was elected 
as a Democrat\ldots
&
\textbf{Q:} Where was \textbf{James Paul Maher} born? \newline
\textbf{D:} Born in \colorbox{cyan!20}{\textbf{Brooklyn, New York}}, 
\colorbox{yellow!30}{\textbf{Maher}} graduated from St.\ Patrick's 
Academy in Brooklyn. Apprenticed to the hatter's trade, he moved 
to Danbury, Connecticut in 1887 and was employed as a journeyman 
hatter\ldots
\\
\bottomrule
\end{tabular}
\caption{Examples of bias-amplified document pairs from the ColDER 
benchmark~\cite{fayyaz-etal-2025-collapse}, adapted from their Table~1.
Each row shows a query with two documents: $D_1$ amplifies a 
specific bias-inducing characteristic while $D_2$ serves as a 
matched control.
\colorbox{yellow!30}{Head entities} and 
\colorbox{cyan!20}{tail entities / answers} are highlighted.
In all cases, both documents contain the correct answer; they 
differ only in the targeted surface-level feature.}
\label{tab:bias_examples}
\end{table*}

\subsection{Bias Types}

We evaluate four systematic biases identified by \citet{fayyaz-etal-2025-collapse} in dense retrievers:
\begin{enumerate}\setlength{\itemsep}{0pt}
    \item[(i)] \textbf{Brevity Bias:} Retrievers prefer shorter documents over longer ones containing equivalent evidence. Document $D_1$ contains the same factual information as $D_2$ but with significantly fewer tokens.
    \item[(ii)] \textbf{Literal Bias:} Retrievers prioritize exact lexical matches over semantic equivalence. Document $D_1$ uses identical or highly overlapping terminology with the query, while $D_2$ expresses the same information using paraphrases or synonyms.
    \item[(iii)] \textbf{Position Bias:} Retrievers favor content appearing early in documents over information at later positions. In document pairs, $D_1$ places the relevant information at the beginning while $D_2$ positions it toward the end.
    \item[(iv)] \textbf{Repetition Bias:} Retrievers prefer documents with repeated entity mentions. Document $D_1$ contains multiple mentions of key entities, while $D_2$ mentions them fewer times despite containing equivalent factual content.
\end{enumerate}

We refer the reader to \citet{fayyaz-etal-2025-collapse} for rigorous definitions of these bias types, and to \autoref{tab:bias_examples} for examples of each of these biases.

\subsection{Dense Retrievers}

We evaluate six state-of-the-art dense retrievers representing diverse architectural approaches:

\noindent\textbf{(1) Contrastive Bi-Encoders:} Contriever-MSMARCO~\cite{izacard2021unsupervised}, COCO-DR Base MSMARCO~\cite{yu2022coco}, Dragon RoBERTa~\cite{lin2023train}, DRAGON+~\cite{lin2023train} 

\noindent\textbf{(2) Generative Pretraining:} RetroMAE MSMARCO FT~\cite{xiao-etal-2022-retromae} 

\noindent\textbf{(3) Token-Level Interaction:} ColBERTv2 \cite{santhanam-etal-2022-colbertv2} 

This diverse set represents the major paradigms in modern dense retrieval---contrastive learning, diverse data augmentation, and late interaction architectures---enabling us to assess whether query rewriting effects on biases are architecture-specific or systemic across different retrieval approaches.

\subsection{Query Enhancement Techniques}

We evaluate four query enhancement approaches that represent the spectrum of techniques used in modern RAG systems. 

\noindent\textbf{(1) LLM-based Query Rewriting:} We use an LLM to reformulate the original query for better semantic matching. Given query $q$, we prompt the model to produce the rewritten query $q_{\text{rewrite}}$. 

\noindent\textbf{(2) HyDE:} Following \citet{gao-etal-2023-precise}, we leverage the Hypothetical Document Embeddings (HyDE) approach, which uses an LLM to generate pseudo-documents that might hypothetically contain the answer to a query rather than using the query directly for retrieval. For a given query $q$, we prompt a language model $\mathcal{M}$ to generate a pseudo-document $\hat{d} = \mathcal{M}(\text{prompt}(q))$ 
where $\text{prompt}(q)$ instructs the model to generate a passage that would contain the answer to query $q$. The generated hypothetical document $\hat{d}$ is then embedded and used as the query representation for retrieval. The key insight of HyDE is that hypothetical documents often provide richer semantic context than short queries, potentially capturing vocabulary, style, and relational patterns that better align with actual target documents. 

\noindent\textbf{(3) Query2Doc (Q2D):} We adopt the Query2Doc approach introduced by \citet{wang-etal-2023-query2doc}, which augments queries with LLM-generated pseudo-documents rather than replacing them entirely. For a given query $q$, we prompt a language model $\mathcal{M}$ to generate a pseudo-document $\hat{d}$ that might be relevant to the query. This generated passage is then concatenated with the original query to form an expanded query representation $\hat{q}_{\text{Q2D}} = q \oplus \hat{d}$ 
where $\oplus$ denotes concatenation. The key distinction from HyDE is that Query2Doc preserves the original query signal while enriching it with generated context. Since our retrieval scenario is inherently zero-shot, we do not utilize few-shot exemplars for generation of pseudo-documents like the original Query2Doc paradigm. This augmented query $q_{\text{Q2D}}$ is then embedded and used for retrieval.

\medskip

In line with recent efforts toward domain-specific LLMs in RAG systems~\cite{zhang2024raft}, we test whether domain adaptation can improve query rewriting for bias mitigation. Specifically, we continually pretrain $\mathcal{M}$ on the original Re-DocRED document for a given document pair from ColDeR, before applying HyDE or Query2Doc. Specifically, we use Low-Rank Adaptation (LoRA)~\cite{hu2022lora} to adapt the model to document $d$ by minimizing: 
\begin{align*} 
\mathcal{L}_{\text{CPT}} = -\sum_{i=1}^{|d|} \log P_\theta(w_i \mid w_{<i}, d) 
\end{align*} 
where $\theta$ represents the LoRA parameters and $w_i$ denotes the $i^{\text{th}}$ token in document $d$. The hypothesis is that this adapted model $\mathcal{M}_{\text{domain}}$ familiar with domain-specific vocabulary and factual patterns may generate pseudo-documents that better align with actual evidence, potentially reducing reliance on surface-level features. We use LoRA with rank $r=8, \alpha=16$, and train for $10$ epochs per document. This gives us our final two techniques:

\noindent\textbf{(4) HyDE with Continual Pretraining (HyDE-CPT):} We use the domain-adapted model $\mathcal{M}_{\text{domain}}$ to use to generate psuedo-document $\hat{d}_{\text{CPT}} = \mathcal{M}_{\text{domain}}(\text{prompt}(q))$ 
following the HyDE prompt, and then use the embedding of $\hat{d}_{\text{CPT}}$ for retrieval.

\noindent\textbf{(5) Query2Doc with Continual Pretraining (Q2D-CPT):} We concatenate the original query to the pseudo-document generated by the domain-adapted model $\mathcal{M}_{\text{domain}}$ to obtain $\hat{q}_{\text{Q2D-CPT}} = q \oplus \hat{d}_{\text{CPT}}$ 
and use its embedding for retrieval.
\medskip

To ensure that the effects we observe are robust across the LLM used for query enhancement, we test open-source models from two family of models: Gemma 3~\cite{team2025gemma} (\texttt{google/gemma-3-12b-it}) and Qwen3~\cite{yang2025qwen3} (\texttt{Qwen/Qwen3-4B-Instruct-2507}).

\subsection{Evaluation Metrics}

For each bias type and query enhancement technique, we measure retrieval preferences using paired t-tests~\cite{ross2017paired} on score differences $\Delta = [\delta_1, \ldots, \delta_N]$ across all query-document pairs. We report \emph{mean $|t|$-statistic}, the absolute value of the t-statistic measuring the strength of bias, as well as the percentage decrease in mean $|t|$-statistic compared to the vanilla query. Furthermore, we compute $p-$values to check for statistical significance of the bias at $\alpha = 0.05$ with Bonferroni correction~\cite{bonferroni1935calcolo}. Lower $|t|$-statistic values indicate reduced bias, with values approaching zero suggesting factual grounding rather than reliance on surface characteristics.

%% file: latex/4Results.tex
\section{Results}\label{sec:results}

We now present results for query enhancements using Gemma-3-12B-IT, and defer the results for Qwen3-4B-Instruct to \autoref{app:qwen3-results} since we observe very similar trends overall.

\subsection{Overall Trend} 

\begin{table}[t]
\centering
\sffamily
\resizebox{\columnwidth}{!}{
\begin{tabular}{lcccc}
\textbf{Method} & \textbf{Mean $|t|$} & \textbf{Sig. Biases} & \textbf{Reduction} \\
\midrule
\cellcolor{badcell}Baseline & \cellcolor{badcell}8.72 $\pm$ 5.32 & \cellcolor{badcell}21/24 & \cellcolor{badcell}-- \\
\hdashline
\cellcolor{goodcell}Rewrite & \cellcolor{goodcell}4.02 $\pm$ 2.17 & \cellcolor{goodcell}13/24 & \cellcolor{goodcell}+53.9\% \\
\hdashline
HyDE & 6.95 $\pm$ 4.73 & 20/24 & +20.3\% \\
HyDE-CPT & 6.78 $\pm$ 4.79 & 19/24 & +22.3\% \\
\hdashline
Q2D & 6.15 $\pm$ 5.96 & 16/24 & +29.5\% \\
Q2D-CPT & 6.07 $\pm$ 5.72 & 17/24 & +30.4\% \\
\bottomrule
\end{tabular}}
\caption{Summary of retrieval bias across query enhancement methods. Mean $|t|$-statistic averaged across all retrievers and bias types. Sig. Biases: number of retriever-bias combinations showing significant bias ($p < 0.05$ after Bonferroni correction). Reduction: percentage decrease in $|t|$ compared to Baseline.\vspace{-12pt}}
\label{tab:bias-summary}
\end{table}

\autoref{tab:bias-summary} summarizes retrieval bias across all query enhancement methods, aggregated over six retrievers and four bias types (24 total retriever-bias combinations). All query enhancement techniques reduce bias relative to the baseline, though the magnitude varies substantially. Most strikingly, simple LLM-based query rewriting achieves the largest reduction (53.9\%), cutting the mean $|t|$-statistic from 8.72 to 4.02 and reducing the number of statistically significant biases from 21 to 13 out of 24 combinations. The pseudo-document generation methods---HyDE and Query2Doc---provide more modest improvements, with reductions ranging from 20.3\% to 30.4\%. Query2Doc variants outperform HyDE variants, suggesting that preserving the original query signal while augmenting with generated content is more effective than replacing the query entirely. Continual pretraining (CPT) offers marginal gains over the base methods (HyDE-CPT: +2.0\% over HyDE; Q2D-CPT: +0.9\% over Q2D), indicating that domain adaptation primarily improves factual grounding rather than bias mitigation. However, these aggregate statistics mask important variation across bias types that we examine in detail below.

\subsection{Reduction in Retrieval Bias}

\begin{figure}[t]
    \centering
    \includegraphics[width=\linewidth]{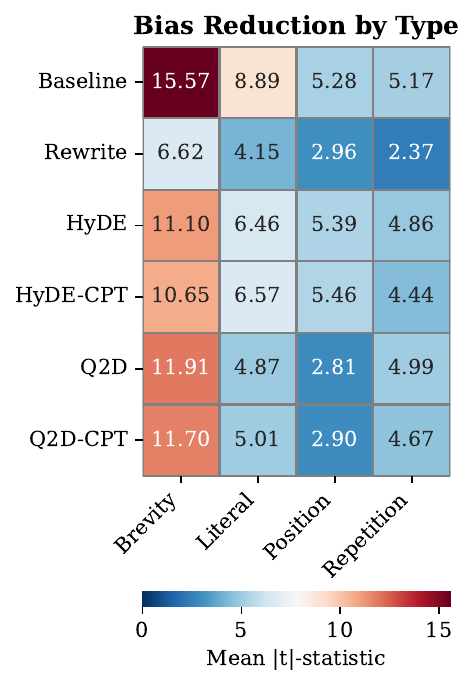}
    \caption{\textbf{Mean $|t|$-statistic by bias type and method (averaged across 6 retrievers).} Lower values indicate reduced bias. All $|t|$ values were averaged irrespective of statistical significance.\vspace{-12pt}}
    \label{fig:bias-across-retrievers}
\end{figure}

\autoref{fig:bias-across-retrievers} presents the mean $|t|$-statistic across bias types and query enhancement methods, averaged over all retrievers. We observe several patterns:

\paragraph{Simple rewriting is surprisingly effective.} Contrary to our intuition, simple LLM-based query rewriting achieves the strongest overall bias reduction, decreasing the average $|t|$-statistic from 8.72 to 4.02. This simple baseline outperforms more sophisticated pseudo-document generation methods across all four bias types. The result suggests that much of the bias in dense retrieval may stem from the surface-level characteristics of user queries themselves---their brevity, specific lexical choices, or syntactic patterns---which even basic reformulation can address.

\paragraph{Pseudo-document methods show bias-specific trade-offs.} HyDE and Query2Doc show notably different bias reduction profiles. HyDE reduces brevity bias substantially (15.57 $\to$ 11.10) but slightly \textit{exacerbates} position bias (5.28 $\to$ 5.39). We hypothesize that HyDE-generated documents, which replace the query entirely, may introduce their own positional patterns that interact adversely with document encodings. In contrast, Query2Doc—which preserves the original query while augmenting it—achieves the strongest position bias reduction of any method (5.28 $\to$ 2.81, a 47\% decrease). This suggests that retaining the original query signal provides an anchoring effect that prevents the introduction of new biases.

\paragraph{Brevity bias remains the most severe.} Across all methods, brevity bias exhibits the highest $|t|$-statistics, indicating that dense retrievers' preference for shorter documents is particularly robust to query-side interventions. Even the best-performing method (\emph{Rewrite}) leaves a residual bias of 6.62, compared to near-complete mitigation for position bias (2.81 with \emph{Q2D}). This asymmetry suggests that brevity bias may be more deeply encoded in document representations, requiring retriever-level interventions rather than query modifications alone.

\paragraph{Literal matching responds well to most techniques.} All query enhancement techniques substantially reduce literal matching bias, with an average reduction of 39.12\%. This suggests that these techniques dilute exact lexical overlap between short queries and documents, shifting retrieval from lexical toward semantic matching.

\paragraph{Repetition bias proves resistant to pseudo-document methods.}  While simple rewriting cuts repetition bias by 54\% (5.17 $\to$ 2.37), HyDE and Query2Doc show minimal improvement (5.17 $\to$ 4.86 and 4.99, respectively). This asymmetry suggests that LLM-generated pseudo-documents may inadvertently repeat query terms---a natural consequence of generating text conditioned on answering the query---thereby perpetuating rather than mitigating repetition bias. See \autoref{app:generated-query-examples} for examples of generated pseudo-documents which confirms the same.

\subsection{Variation Across Dense Retrievers}

\begin{figure}[t]
    \centering
    \includegraphics[width=\linewidth]{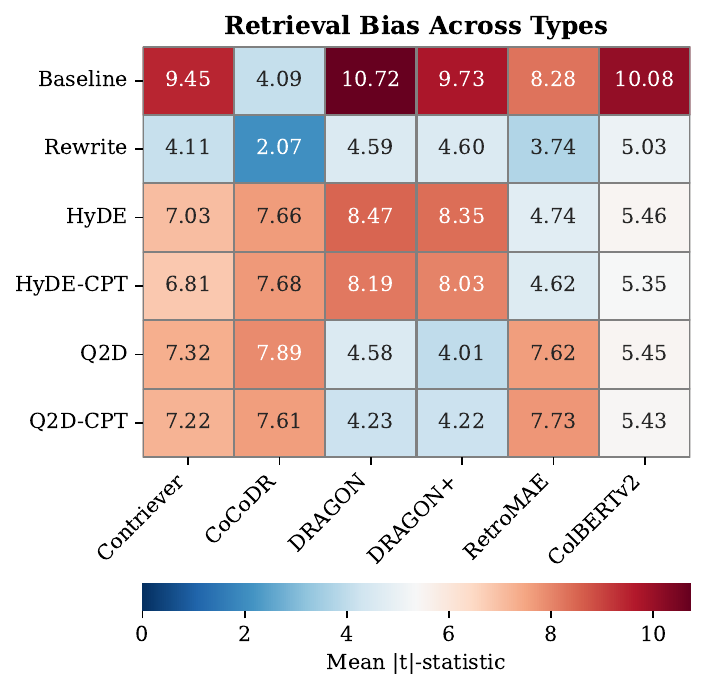}
    \caption{\textbf{Mean $|t|$-statistic by retriever and query enhancement method, averaged across all four bias types.} Lower values indicate reduced bias. Simple rewriting consistently reduces bias, while pseudo-document methods show differential effects.\vspace{-12pt}}
    \label{fig:bias-across-types}
\end{figure}

\autoref{fig:bias-across-types} reveals the heterogeneity in how different retrievers respond to various query enhancement techniques.

\paragraph{Retriever-specific effects.} We observe that simple query rewriting achieves consistent bias reduction across all six retrievers, with mean $|t|$-statistics ranging from 2.07 (CoCoDR) to 5.03 (ColBERTv2). This consistency stands in stark contrast to the pseudo-document methods, which exhibit highly retriever-dependent effects. For example, Q2D dramatically reduces bias for DRAGON and DRAGON+ (10.72 $\to$ 4.58 and 9.73 $\to$ 4.01, respectively), yet \textit{increases} bias for CoCoDR (4.09 $\to$ 7.89) and RetroMAE (8.28 $\to$ 7.62). Similarly, HyDE substantially harms CoCoDR (4.09 $\to$ 7.66), nearly doubling its bias level. These findings suggest that the interaction between pseudo-document generation and retriever architecture is non-trivial and potentially unpredictable.

\paragraph{Architectural patterns.} Several architecture-specific trends emerge from these results. The DRAGON variants---both trained with diverse augmentation strategies---respond exceptionally well to Query2Doc, achieving bias levels comparable to simple rewriting (4.01--4.58 vs. 4.59--4.60). We hypothesize that these retrievers, having been exposed to diverse query-document pairings during training, are better equipped to leverage the augmented context that Query2Doc provides. In contrast, CoCoDR exhibits anomalous behavior: despite having the lowest baseline bias (4.09), it is the only retriever for which all pseudo-document methods increase bias. This may reflect CoCoDR's continuous contrastive distillation objective, which could make it sensitive to the distributional shift introduced by LLM-generated content. ColBERTv2's late interaction architecture yields remarkably stable results across all enhancement methods (5.03--5.46), suggesting that token-level matching may be inherently more robust to query formulation changes than single-vector representations. 

\subsection{Effect of Query Enhancement on Interplay Between Bias Types}
\label{sec:foil}

While Section~\ref{sec:results} demonstrated that query rewriting techniques reduce individual bias types, a natural question arises: \textit{do these improvements transfer to scenarios where multiple biases act concurrently?} 

To investigate this, we evaluate on the \textsc{Foil} subset of ColDeR, which presents an adversarial set of $250$ document pairs where document $D_1$ exploits multiple bias-inducing features (brevity, answer position, lexical overlap, repetition) while lacking the answer, and $D_2$ contains the correct evidence embedded within unrelated context. We report results for RetroMAE, Contriever, and CoCoDR here, to capture the spectrum of performances.

\begin{table}[t]
\centering
\sffamily
\resizebox{\columnwidth}{!}{
\begin{tabular}{lccc}
\textbf{Method} & \textbf{RetroMAE} & \textbf{Contriever} & \textbf{CoCoDR} \\
\midrule
\rowcollight Baseline & 0.0\% & 0.8\% & 2.0\% \\
\hdashline
Rewrite & \cellcolor{badcell}0.0\% & \cellcolor{badcell}0.8\% & \cellcolor{badcell}3.6\% \\
\hdashline
HyDE & 2.8\% & 4.0\% & 20.0\% \\
HyDE-CPT & 2.8\% & \cellcolor{goodcell}5.2\% & \cellcolor{goodcell}26.0\% \\
\hdashline
Q2D & 2.8\% & 3.2\% & 18.0\% \\
Q2D-CPT & \cellcolor{goodcell}3.2\% & 3.6\% & 19.2\% \\
\midrule
\textit{Best improv.} & --- & 6.5$\times$ & 13$\times$ \\
\bottomrule
\end{tabular}}
\caption{\textbf{\textsc{Foil} dataset accuracy across retrievers.} CoCoDR shows the strongest response to query enhancement, with HyDE-CPT achieving 26\% accuracy (13$\times$ over baseline). Simple rewriting provides minimal improvement across all retrievers. \vspace{-12pt}}
\label{tab:foil-comparison}
\end{table}

From \autoref{tab:foil-comparison}, we see that simple query rewriting provides minimal improvement on \textsc{Foil} despite achieving the largest reduction in individual bias metrics (Section~\ref{sec:results}). In contrast, pseudo-document generation methods show substantial gains---on CoCoDR, HyDE achieves 10$\times$ improvement (20.0\%) and Query2Doc achieves 9$\times$ improvement (18.0\%) over baseline. These methods fundamentally alter the query representation through semantic expansion and length normalization, thereby providing more robust resistance to adversarial bias combinations. CPT provides additional benefit over vanilla techniques, with HyDE-CPT achieving the strongest performance at 26.0\% (13$\times$ over baseline), which suggests that document-specific fine-tuning addresses not just relevance matching but also enhances bias robustness.

These findings motivate a deeper investigation: \emph{what distinguishes methods that genuinely reduce bias sensitivity from those that merely obscure it?} We address this question through a deeper mechanistic analysis in the following section.

\section{Mechanistic Analysis of Bias Reduction}
\label{sec:analysis}

We now investigate why query-enhancement techniques might succeed or fail at mitigating biases. We conduct an analysis to measure how rewriting affects the correlation between retrieval scores and bias-inducing features.

If a retriever exhibits bias toward a particular document feature (e.g., shorter length, answer position), we expect retrieval scores to correlate with that feature independent of relevance. Our intuition is that effective bias mitigation should reduce this correlation, and we operationalize this intuition by computing the Spearman correlation $\rho$ between retrieval scores and bias-inducing features across the ColDeR benchmark.

For each query $q$ (original or rewritten) and document $d$, we compute the retrieval score $\mathcal{S}(q, d)$ and extract the corresponding bias feature $f(q, d)$:
\begin{enumerate}
    \item[(i)] \textbf{Brevity bias:} Document length in tokens, $f(q, d) = |d|$.
    
    \item[(ii)] \textbf{Literal matching bias:} Jaccard similarity~\cite{jaccard1901etude} between query and document term sets. Let $\mathcal{T}(q)$ and $\mathcal{T}(d)$ denote the sets of unique terms in query $q$ and document $d$, respectively. Then, we define:
    \begin{align*}
        f(q, d) = \frac{|\mathcal{T}(q) \cap \mathcal{T}(d)|}{|\mathcal{T}(q) \cup \mathcal{T}(d)|}
    \end{align*}
    
    \item[(iii)] \textbf{Position bias:} Normalized answer position within the document, where $\text{pos}(a, d)$ denotes the character offset of answer span $a$ in document $d$:
    \begin{align*}
        f(q, d) = \frac{\text{pos}(a, d)}{|d|}  \in [0, 1]
    \end{align*}
    
    \item[(iv)] \textbf{Repetition bias:} Average term frequency of query terms within the document. Let $\text{tf}(t, d)$ denote the number of occurrences of term $t$ in document $d$:
    \begin{align*}
        f(q, d) = \frac{1}{|\mathcal{T}(q)|} \sum_{t \in \mathcal{T}(q)} \text{tf}(t, d)
    \end{align*}
\end{enumerate}

We then compute $\rho_c = \text{Spearman}(\mathcal{S}_c, f)$ for each query condition $c$. A retriever with no systematic bias would yield $\rho \approx 0$, while higher absolute correlations indicate stronger bias. We quantify bias reduction as:
\begin{align}
    \Delta\rho = |\rho_{\text{original}}| - |\rho_{\text{rewritten}}|
\end{align}
\noindent where $\Delta\rho > 0$ indicates that rewriting reduced the retriever's sensitivity to the bias-inducing feature.

\paragraph{Findings:} \autoref{tab:decorrelation} presents the mean correlation reduction $\Delta\rho$ across six dense retrievers. We observe several key patterns:

\begin{table}[t]
\centering
\sffamily
\resizebox{\columnwidth}{!}{
\begin{tabular}{lcccc}
\textbf{Method} & \textbf{Brevity} & \textbf{Literal} & \textbf{Position} & \textbf{Repetition} \\
\midrule
\rowcollight \textit{Baseline $|\rho_q|$} & 0.36 & 0.43 & 0.11 & 0.21 \\
\hdashline
Rewrite   & \cellcolor{goodcell}+10\% & \cellcolor{badcell}$-$7\%  & \cellcolor{badcell}$-$373\% & \cellcolor{badcell}$-$112\% \\
\hdashline
HyDE      & \cellcolor{goodcell}+56\% & \cellcolor{goodcell}+20\% & \cellcolor{badcell}$-$2\%   & \cellcolor{badcell}$-$59\% \\
HyDE-CPT  & \cellcolor{goodcell}+49\% & \cellcolor{goodcell}+19\% & \cellcolor{goodcell}+54\%  & \cellcolor{badcell}$-$9\% \\
\hdashline
Q2D       & \cellcolor{goodcell}+48\% & \cellcolor{goodcell}+36\% & \cellcolor{badcell}$-$55\%  & \cellcolor{badcell}$-$13\% \\
Q2D-CPT   & \cellcolor{goodcell}+53\% & \cellcolor{goodcell}+32\% & \cellcolor{goodcell}+23\%  & \cellcolor{badcell}$-$21\% \\
\bottomrule
\end{tabular}}
\caption{\textbf{Feature-score decorrelation analysis.} $|\rho_q|$ represents baseline Spearman correlation for query $q$. Values show percentage reduction in $|\rho|$ after rewriting query $q$ (\colorbox{goodcell}{positive} = reduced sensitivity, \colorbox{badcell}{negative} = increased sensitivity). Results averaged across all 6 dense retrievers studied.\vspace{-12pt}}
\label{tab:decorrelation}
\end{table}

\paragraph{Brevity bias.} All rewriting methods substantially reduce correlation with document length. Pseudo-document methods achieve the largest gains (HyDE: +56\%, Q2D: +48\%), consistent with our hypothesis that generating longer text shifts the query representation away from favoring brief documents. This finding aligns with Section~\ref{sec:results}, where all methods reduced brevity bias.

\noindent\textbf{Literal matching bias.} Query2Doc variants achieve substantially larger reductions in literal bias compared to HyDE. We attribute this to a key mechanistic difference: Query2Doc preserves original query terms while adding semantic context, effectively \textit{diluting} rather than \textit{replacing} lexical signals. HyDE's complete query replacement may introduce new lexical artifacts from the generated pseudo-document. Notably, simple rewriting slightly \textit{increases} literal bias ($-$7\%), suggesting that paraphrasing without content expansion actually reinforces lexical overlap as the rewritten query may use more ``retrieval-friendly'' vocabulary.

\noindent\textbf{Repetition bias.} All approaches \textit{increase} sensitivity to repetition, with simple rewriting showing the most severe degradation. This represents a systematic failure that explains the resistance of repetition bias to pseudo-document methods. Analysis of generated pseudo-documents shows that LLMs naturally produce repetitive text patterns when answering queries, which correlates with documents containing repeated query terms. The $|t|$-statistic improvements likely reflect increased score variance rather than true debiasing.

\noindent\textbf{Position bias.} Position bias reveals the strongest mechanistic differences. Simple rewriting greatly position sensitivity, yet achieved strong $|t|$-statistic reduction in Section~\ref{sec:results}. This paradox suggests that simple rewriting reduces position bias through increased retrieval noise rather than principled decorrelation. In contrast, HyDE-CPT achieves genuine decorrelation. Only Q2D-CPT combines strong $|t|$-statistic reduction with positive decorrelation, suggesting that continual pretraining is essential for robust position bias mitigation.

\paragraph{Reconciling $|t|$-statistics and decorrelation:} These findings reveal a crucial insight that \emph{reducing overall bias (as measured by $|t|$-statistics) does not imply reducing sensitivity to bias-inducing features}. We identify two distinct mechanisms:

\noindent\textbf{(i) Variance-based reduction:} Simple rewriting achieves strong $|t|$-statistic improvements while \textit{increasing} feature-score correlations for position and repetition biases. The mechanism appears to be increased retrieval score variance---noisier scores reduce the statistical power to detect systematic preferences, lowering $|t|$-statistics without addressing underlying bias sensitivity. This also explains why simple rewriting fails entirely on \textsc{Foil} (\autoref{tab:foil-comparison}); when biases act together, the underlying sensitivities may compound further.

\noindent\textbf{(ii) Decorrelation-based reduction:} Pseudo-document methods, particularly the CPT variants, achieve bias reduction through principled decorrelation from bias-inducing features. Specifically, HyDE-CPT and Q2D-CPT show positive decorrelation for brevity, literal matching, \textit{and} position biases simultaneously. This more robust mechanism explains their gains on \textsc{FOIL}, where genuine insensitivity to bias features provides resistance to adversarial combinations.

To further understand why simple rewriting shows contrasting effects, we analyzed the linguistic transformations applied by the LLM-based rewriting (\autoref{tab:query_analysis} in \autoref{app:analysis-query-rewriting}). Our key observation is that these transformations are predominantly \textit{syntactic rather than semantic}, introducing sufficient variance to reduce the paired t-statistic without actually reducing the retriever's sensitivity to bias-inducing features.

%% file: latex/5Discussion.tex
\section{Discussion}

\noindent\textbf{Implications for researchers.} Our findings suggest a taxonomy of retrieval biases based on their responsiveness to query-side interventions. \textit{(i) Query-document interaction biases}---such as literal matching---arise from how queries and documents are compared and can be mitigated through query transformation. \textit{(ii) Document encoding biases}---particularly position bias---appear embedded in document representations and persist regardless of query formulation, likely requiring retriever-level modifications. Additionally, our decorrelation analysis reveals that aggregate metrics can obscure different underlying mechanisms, and we recommend future work also report feature-score correlations. Finally, the increase in repetition sensitivity across all methods represents a systematic failure mode that warrants further investigation.

\noindent\textbf{Implications for practitioners.} When selecting query enhancement techniques, practitioners should consider specific bias vulnerabilities rather than aggregate metrics. Simple rewriting achieves strong overall reduction but fails when multiple biases act together. Critically, enhancement effects are also retriever-dependent, so techniques should be validated on specific retrievers before deployment. For high-stakes domains, we recommend CPT variants despite computational overhead, as our findings reveal that decorrelation-based reduction generalizes better to adversarial conditions than variance-based reduction.

%% file: latex/6Conclusion.tex
\section{Conclusion}

In this work, we present the first systematic study of how query enhancement techniques affect dense retrieval biases in RAG systems. Our evaluation reveals that while all techniques reduce aggregate bias, they operate through fundamentally different mechanisms. Simple rewriting achieves strong overall reduction but increases sensitivity to bias-inducing features; pseudo-document methods with continual pretraining achieve more robust improvements through genuine decorrelation. We establish a taxonomy distinguishing query-document interaction biases, which yield to query-side interventions, from document encoding biases, which likely require retriever-level modifications. Our findings provide practical guidance for deploying bias-aware RAG systems, and highlight the continued need for retrieval paradigms that prioritize semantic relevance over superficial characteristics.

%% file: latex/7Limitations.tex
\section{Limitations}

\paragraph{(1) Benchmark and retriever scope:} Our evaluation relies on the ColDeR benchmark, which provides controlled document pairs per bias type derived from Re-DocRED. ColDeR, to the best of our knowledge, is currently the only openly  available resource enabling this kind of controlled, multi-dimensional bias evaluation. Our statistical analyses (paired $t$-tests with Bonferroni correction) are appropriately powered for this sample size, and we observe consistent effect directions across six retrievers, increasing confidence in the findings' reliability. Nevertheless, we acknowledge that these controlled pairs may not fully capture the distribution of biases in real-world retrieval scenarios where multiple subtle biases co-occur in less structured ways. Additionally, we evaluate only dense bi-encoder and late-interaction retrievers, primarily because of their widespread utility and effectiveness today. Our findings may not generalize to sparse retrievers (e.g., BM25), hybrid systems, or emerging retrieval paradigms such as generative retrieval.

\paragraph{(2) Query enhancement coverage:} We evaluate five query enhancement techniques using two open-source LLMs (Gemma-3 and Qwen3). Other methods such as Step-Back Prompting, multi-query fusion, or chain-of-thought augmented retrieval may exhibit different bias profiles. However, our key contribution is not tied to a specific query-enhancement type. Furthermore, proprietary LLMs with stronger instruction-following capabilities might produce pseudo-documents with different characteristics, potentially affecting bias outcomes, which we could not evaluate in this work due to cost constraints.

\paragraph{(3) Downstream evaluation:} Our analysis focuses on retrieval-stage biases measured through score differences and correlations. We do not evaluate how these biases propagate to downstream generation quality in end-to-end RAG systems. A retrieval bias that appears significant in isolation may be attenuated, or amplified, by the downstream language model, and understanding this interaction remains an important direction for future work.

%% file: latex/Acknowledgments.tex


%% file: latex/EthicsStatement.tex
\section*{Ethical Considerations}

Our work aims to understand and mitigate biases in dense retrieval systems, which we view as a net positive for the responsible and fair deployment of RAG systems. However, we acknowledge that detailed characterization of retrieval biases could theoretically be misused to craft adversarial documents that exploit these biases to manipulate retrieval rankings. However, we believe the benefits of transparency outweigh this risk, as awareness of these biases enables practitioners to implement appropriate safeguards.

More broadly, the biases we study have implications for fairness in information access for users of modern generative search systems. For example, brevity bias may systematically disadvantage comprehensive sources, and literal matching bias may favor keyword-filled content over semantically rich sources. These effects are particularly concerning in high-stakes domains such as legal, sociotechnical, and medical information retrieval, where biased retrieval could lead to incomplete or skewed information being surfaced to end users~\cite{liu2023evaluating}. Our findings therefore underscore the importance of bias-aware evaluation in retrieval system development.

%% file: latex/Appendix.tex




\section{Examples of Generated Query Enhancements}\label{app:generated-query-examples}

In this section we provide examples of query enhancements across methods.

\begin{tcolorbox}[
  title=\textbf{Vanilla Queries},
  colback=black!10,
  colframe=black!60!black,
  fonttitle=\bfseries,
  fontupper=\sffamily\small, 
  rounded corners,
  breakable 
]
1. When was The Private Life of Helen of Troy published?\\

2. Where is Los Angeles located in or next to a body of water?\\

3. Where was Cesare Mori born?
\end{tcolorbox}

\begin{tcolorbox}[
  title=\textbf{LLM Rewritten Queries},
  colback=black!10,
  colframe=black!60!black,
  fonttitle=\bfseries,
  fontupper=\sffamily\small, 
  rounded corners,
  breakable 
]
1. What is the publication date of The Private Life of Helen of Troy?\\

2. What body of water is Los Angeles located near or in?\\

3. What is the birthplace of Cesare Mori?
\end{tcolorbox}

\begin{tcolorbox}[
  title=\textbf{HyDE Queries},
  colback=black!10,
  colframe=black!60!black,
  fonttitle=\bfseries,
  fontupper=\sffamily\small, 
  rounded corners,
  breakable 
]
1. The ``Private Life of Helen of Troy'' was published in 1920 by the renowned author, Robert Graves. This groundbreaking work, a reimagining of the Trojan War legend, offered a fresh perspective on Helen's character and the events surrounding the war.\\

2. Los Angeles is located on the southern coast of California, a state situated on the Pacific Ocean. The city is situated on the shores of the Pacific Ocean, with the Pacific Ocean bordering its western side. The city is also bordered by the Los Angeles Basin, a large inland basin that is partially surrounded by mountains.\\

3. Cesare Borgia, the Italian Renaissance figure, died in Rome, Italy on 18th of January 1507.  He was a powerful and controversial figure who served as Cardinal and later as the Duke of Valentinois.  His death was a significant event in the history of the Borgia family and the Italian Renaissance.
\end{tcolorbox}

\begin{tcolorbox}[
  title=\textbf{Query2Doc Queries},
  colback=black!10,
  colframe=black!60!black,
  fonttitle=\bfseries,
  fontupper=\sffamily\small, 
  rounded corners,
  breakable 
]
1. When was The Private Life of Helen of Troy published?\\ The ``Private Life of Helen of Troy'' was published in 1920 by the renowned author, Robert Graves. This groundbreaking work, a reimagining of the Trojan War legend, offered a fresh perspective on Helen's character and the events surrounding the war.\\

2. Where is Los Angeles located in or next to a body of water?\\ Los Angeles is located on the southern coast of California, a state situated on the Pacific Ocean. The city is situated on the shores of the Pacific Ocean, with the Pacific Ocean bordering its western side. The city is also bordered by the Los Angeles Basin, a large inland basin that is partially surrounded by mountains.\\

3. Where was Cesare Mori born?\\ Cesare Borgia, the Italian Renaissance figure, died in Rome, Italy on 18th of January 1507.  He was a powerful and controversial figure who served as Cardinal and later as the Duke of Valentinois.  His death was a significant event in the history of the Borgia family and the Italian Renaissance.
\end{tcolorbox}

\section{Results for Qwen3}\label{app:qwen3-results}

\begin{figure*}[ht]
    \centering
    \includegraphics[width=0.45\linewidth]{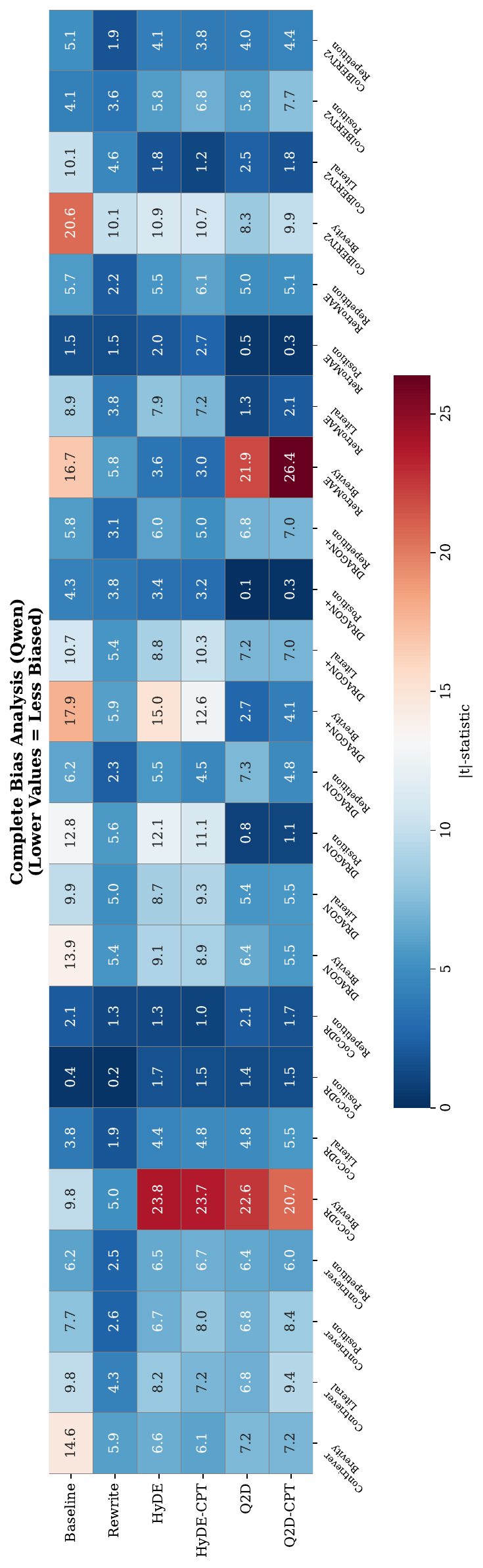}
    \caption{\textbf{Complete bias analysis using Qwen3-4B-Instruct for query enhancement.} Each cell shows the $|t|$-statistic measuring retrieval bias strength across retrievers and biases. Lower values indicate reduced bias. Results demonstrate consistent patterns with Gemma-3-12B-IT, confirming generalizability of our findings.}
    \label{fig:full_heatmap_qwen}
\end{figure*}

To verify that our findings are not specific to the choice of language model used for query enhancement, we replicate our main experiments using Qwen3-4B-Instruct as an alternative to Gemma-3-12B-IT. Figure~\ref{fig:full_heatmap_qwen} presents the complete bias analysis across all retrievers and bias types.

The results show consistent patterns, with simple rewriting achieveing substantial bias reduction on individual metrics, while HyDE and Query2Doc show similar moderate improvements. This consistency across model architectures and scales (4B vs. 12B parameters) suggests that our findings reflect fundamental properties of query transformation strategies rather than artifacts of a specific LLM's generation characteristics. 

\section{Further Analysis of Simple LLM-Based Query Rewriting}\label{app:analysis-query-rewriting}

\begin{table}[h]
\centering
\sffamily
\resizebox{\columnwidth}{!}{
\begin{tabular}{lcc}
\textbf{Metric} & \textbf{Baseline} & \textbf{Rewrite} \\
\midrule
\rowcollight Avg. query length (words) & 7.3 & 8.3 \\
Length change (words) & --- & +1.0 \\
\rowcollight Term preservation (\%) & 100 & 57 \\
New terms introduced & --- & 3.8 \\
\rowcollight Unique vocabulary & 420 & 489 \\
Entity preservation (\%) & 100 & 72.5 \\
\bottomrule
\end{tabular}}
\caption{\textbf{Query transformation characteristics across enhancement methods.} Simple rewriting makes minimal syntactic changes (high term/entity preservation). This explains why rewriting reduces measured bias variance but fails on adversarial \textsc{Foil} examples.}
\label{tab:query_analysis}
\end{table}

\autoref{tab:query_analysis} reveals that simple LLM-based rewriting produces minute changes, with queries increase by only 1 word on average (7.3$\to$8.3 words), preserving 57\% of original terms while introducing 3.8 new terms. The transformations are therefore predominantly \emph{syntactic rather than semantic}. For example, \textit{``When was X published?''} becomes \textit{``What is the publication date of X?''} The key entity names (e.g., \textit{``Lake Ewauna,''} \textit{``Miami Sound Machine''}) remain intact in 72.5\% of cases. 

This explains the paradox we observed in our results in Section~\ref{sec:results} and Section~\ref{sec:analysis}: syntactic reformulation introduces sufficient variance to reduce the paired t-statistic---a measure of \emph{consistency} in bias direction---without actually reducing the retriever's sensitivity to bias-inducing lexical features. Therefore, when multiple biases compound in the \textsc{Foil} setting, this surface-level variance fails to provide robustness.

\section{Compute Resources}

All experiments on open-source models were run
on internal organization GPU servers equipped with 2xNVIDIA H100 and 3xNVIDIA A40.